\documentclass[conference]{IEEEtran}
\IEEEoverridecommandlockouts
\usepackage{cite}
\usepackage{amsmath,amssymb,amsfonts}
\usepackage{algorithmic}
\usepackage{graphicx}
\usepackage{textcomp}
\usepackage{xcolor}
\usepackage{tikz}
\usepackage{pgfplots}
\usepackage{todonotes}
\usepackage{tabularx}

\def\BibTeX{{\rm B\kern-.05em{\sc i\kern-.025em b}\kern-.08em
    T\kern-.1667em\lower.7ex\hbox{E}\kern-.125emX}}
\tikzset{every picture/.style={line width=0.75pt}}
\tikzset{every picture/.style={line width=0.75pt}}
\tikzset{cross/.style={cross out, draw=black, fill=none, minimum size=2*(#1-\pgflinewidth), inner sep=0pt, outer sep=0pt}, cross/.default={2pt}}
\usetikzlibrary{positioning, shapes, trees, automata, calc, fit, automata, shapes.misc, arrows, patterns}
\begin{document}
\title{Impact of channel access and transport mechanisms on QoE in GEO-satellite based LTE backhauling systems}

\author{
\IEEEauthorblockN{Nicolas Kuhn}
\IEEEauthorblockA{\textit{CNES}}
\and
\IEEEauthorblockN{David Fernandes}
\IEEEauthorblockA{\textit{VIVERIS TECHNOLOGIES}}
\and
\IEEEauthorblockN{Emmanuel Dubois}
\IEEEauthorblockA{\textit{CNES}}
\and
\IEEEauthorblockN{David Pradas}
\IEEEauthorblockA{\textit{VIVERIS TECHNOLOGIES}}
}

\maketitle

\begin{abstract}

Backhauling services through satellite systems have doubled between 2012 and
	2018.  There is an increasing demand for this service for which
	satellite systems typically allocate a fixed resource.  This solution
	may not help in optimizing the usage of the scarce satellite resource. 

This study measures the relevance of using dynamic resource allocation
	mechanisms for backhaul services through satellite systems.  The
	satellite system is emulated with OpenSAND, the LTE system with
	Amarisoft and the experiments are orchestrated by OpenBACH.  We
	compare the relevance of applying TCP PEP mechanisms and dynamic resource
	allocations for different traffic services by measuring the QoE for web
	browsing, data transfer and VoIP applications. 

The main conclusions are the following. When the system is congested, PEP and
	layer-2 access mechanisms do not provide significant improvements. When
	the system is not congested, data transfer can be greatly improved
	through protocols and channel access mechanism optimization. Tuning the
	Constant Rate Assignment can help in reducing the cost of the
	resource and provide QoE improvements when the network is not
	loaded. 

\end{abstract}

\begin{IEEEkeywords}
BACKHAUL, SATCOM, PEP, ACCESS
\end{IEEEkeywords}

\section{Introduction}
\label{sec:introduction}

Mobile Network Operators (MNO) have the regular need to optimize their
communication infrastructure in order to better manage congestion, guarantee
the quality of the service and maximize income. The end user demand may not be
located close to the already deployed core network of a MNO and answering to it
may not be economically viable. The deployment of LTE cells and their
connection to the core network with a satellite system has proved to be an efficient 
solution to this issue. 

The satellite backhauling service represented 36,750 sites served by satellite
in 2018, which is twice the amount served in 2012. 
As this service is growing strongly and a
source of significant revenue for operators, it is important to present the
issues related to this use case and the necessary considerations allowing to define
future satellite systems.

\begin{figure}[h]
        \centering
        \includegraphics[width =\linewidth]{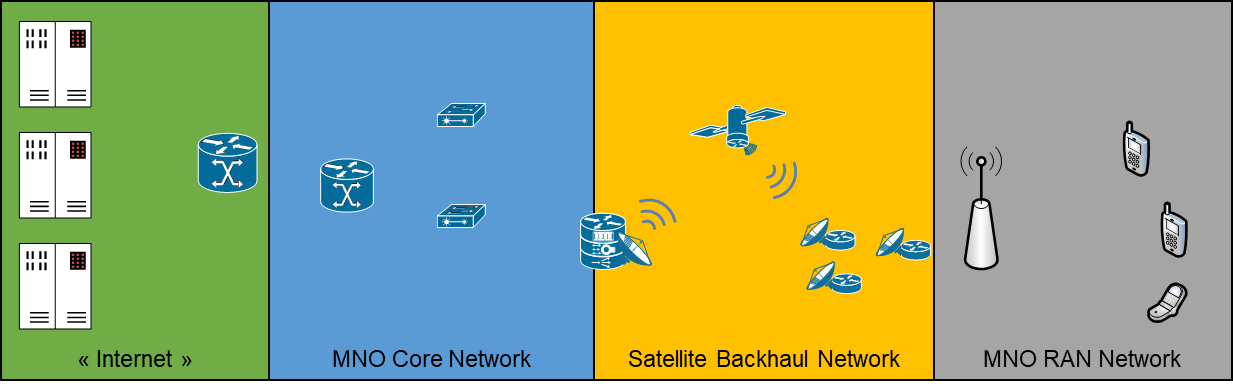}
        \caption{Backhaul services through satellite systems}
        \label{fig:archi-backhaul}
\end{figure}

Figure~\ref{fig:archi-backhaul} shows an architecture of a LTE backhauling service through
satellite. End users (User Equipment, EU) access Radio Access
Network (RAN) of a MNO using LTE standards. The
satellite system connects the RAN to the core network of the MNO.

The backhauling services need for relatively strong end-to-to-end Quality of Service (QoS) requirements. 
In general, variable data rates should be guaranteed by
the QoS mechanisms, the voice from 64~kbps and video from 256~kbps should
also be guaranteed.  There may also be strong requirements on the round trip
time, jitter and packet loss rate. 
\begin{itemize}
	\item[$\blacktriangleright$] GEO satellite backhauling accesses are facing a trade-off between quality of network service, quality of the user experience and the price of the access.
\end{itemize}

Satellite systems may allocate Constant Rate Assignment (CRA) for backhauling
services (as opposed to Rate-Based Dynamic Capacity, RBDC). With CRA, a portion of the
satellite resource is dedicated to a user even if it does not actually use it.
The reduction of the CRA would let the satellite resource management mechanisms
to allocate the unused resource to the systems that actually need it. However,
decreasing the CRA and increasing the RBDC may reduce the Quality of Experience (QoE) 
due to the request-allocation loop inherent to RBDC mechanisms.

This study measures the relevance of using dynamic resource allocation
mechanisms for backhaul services through satellite systems and their impact on
the QoE.  The satellite system is emulated with OpenSAND~\cite{opensand,opensand-site}, the
LTE system with Amarisoft~\cite{amarisoft}, and the experiments are orchestrated by
OpenBACH~\cite{openbach}.  We compare the relevance of applying
PEP~\cite{RFC3135} mechanisms and dynamic resource allocations when the system
is loaded by measuring the QoE for Web browsing, data transfer and VoIP
applications.

The main conclusions are the following. 

\begin{itemize}
	\item[$\blacktriangleright$] When the system is congested, PEP and layer-2 access mechanisms do not provide significant improvements.
	\item[$\blacktriangleright$] When the system is not congested, data transfer can be greatly improved through TCP optimizations.
	\item[$\blacktriangleright$] Tuning the Constant Rate Assignment can help in reducing the cost of the resource and provide QoE improvments when the network is not loaded.
\end{itemize}

\section{Platform details}
\label{sec:platform}

This sections provides details on the exploited platform.

\subsection{On the need for a controlled emulation}

The exploitation of an emulated platform let us consider
mechanisms and algorithms~\cite{when-emulation} that are close to those implemented in deployed systems. When it comes to considering
QoE measurements, simulations may not map actual protocol
performances~\cite{vtc-trustable}.

However, using different proprietary equipments may result in outputs that are
difficult to undersand and analyse. To cover this issue, we propose the
exploitation of a maximum number of opensource softwares towards reproducible
and controlled tests while considering systems as close as possible to real ones.

\subsection{End-to-end emulated platform}

\begin{figure}[h]
        \centering
        \includegraphics[width =\linewidth]{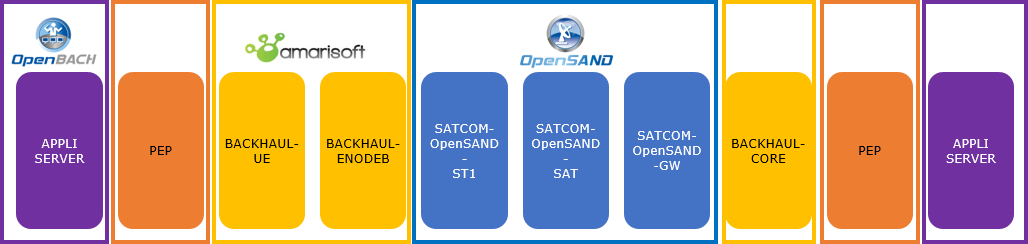}
        \caption{Platform architecture}
        \label{fig:archi-platform}
\end{figure}

Figure~\ref{fig:archi-platform} presents the different equipments that compose
the platform:
\begin{itemize}
	\item Proprietary Performance Enhancing Proxies (PEP);
	\item The cellular network is emulated with AMARISOFT~\cite{amarisoft};
	\item The satellite system is emulated with OpenSAND~\cite{opensand,opensand-site};
	\item The tests are orchestrated by OpenBACH~\cite{openbach}.
\end{itemize}

The PEPs are not deployed within the LTE emulation network since our equipment
could not deal with packets encapsulated within GTP-U tunnels.

\section{Platform validation}
\label{sec:platform}

This sections provides details on how the exploited platform has been validated.

\subsection{Validation strategy}

The experience feedback from previous activities shows that the platform needs to be set up carefully, especially when it integrates so many elements provided by different entities. The following step-by-step procedure has been exploited: 
\begin{itemize}
	\item Prepare the test architecture with the Amarisoft platform and two OpenBACH agents (to emulate the clients);
	\item Launch OpenBACH scenarios allowing an end-to-end QoS analysis and QoE measurements to validate the Amarisoft component;
	\item Add the PEPs in "deactivated" mode (TCP acceleration disabled) at the ends of the system;
	\item Launch OpenBACH scenarios allowing an end-to-end QoS analysis to validate the Amarisoft along with "deactivated" PEP;
	\item Add OpenSAND between the eNodeB and the Core of Amarisoft;
	\item Launch OpenBACH scenarios allowing end-to-end QoS analysis to validate the Amarisoft along with OpenSAND and "deactivated" PEP;
	\item Activate PEP;
	\item Launch OpenBACH scenarios allowing QoS analysis of end-to-end to validate the Amarisoft along with OpenSAND and activated PEP.
\end{itemize}

\subsection{Validation results}

\begin{figure}[h]
        \centering
        \includegraphics[width =\linewidth]{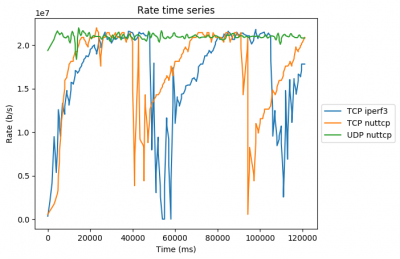}
        \caption{Forward link throughput(b/s)}
        \label{fig:forward_qos_data-rate}
\end{figure}

\begin{figure}[h]
        \centering
        \includegraphics[width =\linewidth]{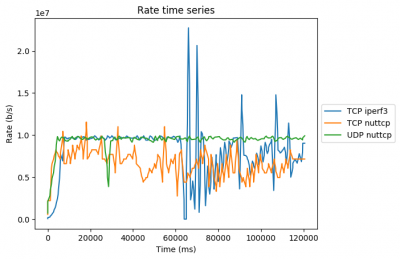}
        \caption{Return link throughput (b/s)}
        \label{fig:return_qos_data-rate}
\end{figure}

\begin{figure}[h]
        \centering
        \includegraphics[width =\linewidth]{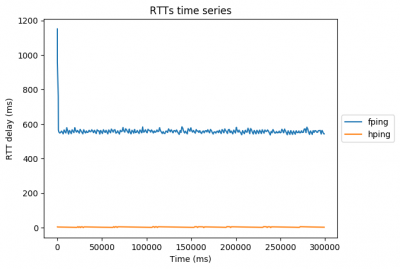}
        \caption{Round Trip Time}
        \label{fig:rtt_qos}
\end{figure}

The results related to the QoS measurements are presented in
Figure~\ref{fig:forward_qos_data-rate}, Figure~\ref{fig:return_qos_data-rate},
Figure~\ref{fig:rtt_qos}.  The traffic has been generated by iperf3 (TCP), nuttcp (TCP/UDP),
fping (ICMP ping) and hping (TCP/IP ping)  The forward channel rate is limited to 20~Mbps and
10~Mbps in the return channel. Changes in TCP throughput already illustrate the
impact of congestion losses on TCP throughput and its ability to use all of the
available capacity. Moreover, the use of OpenBACH makes it possible to obtain
these curves with less effort and greater control and thus assuring the consistency of the
results. The delay measured by hping is $0$ because the traffic is intercepted
by the PEP, which gives an illusion of low latency.

\section{Impact of channel access mechanisms}
\label{sec:access_channel}

This section provides the experiments that have led to assess the impact
of channel access mechanism in a GEO-satellite backhaul system.  The PEP
mechanisms are not activated.

\subsection{Dynamic SATCOM access and uncongested network}

The characteristics of this test scenario are as follows:
\begin{itemize}
	\item Number of connected UEs: 10;
	\item Same application for all UEs: data transfer in download;
	\item All UEs start downloading at the same time; 
	\item OpenSAND limits: 20 Mbps Forward / 1 Mbps in Return;
	\item Flow limitation per UE (SLA) in download: 2 Mbps;
	\item OpenSAND access (total return channel at 1 Mbps):
		\begin{itemize} 
			\item CRA = 50 kbps / RBDC = 1000 kbps
			\item CRA = 100kbps / RBDC = 900 kbps
			\item CRA = 500 kbps / RBDC = 500 kbps
			\item CRA = 1000 kbps /RBDC = 0 kbps
		\end{itemize}
\end{itemize}

The metric reported in this report is the rate convergence time, \textit{i.e.}
the time required for the UE to reach the rate of its SLA. This
enables the analysis of the impact of the access mechanism
on the speed of convergence of congestion control.

\begin{table}[h]
\caption{Rate convergence time when all users start at the beginning of the connection}
\label{table:access_uncongested}
        \begin{tabularx}{\linewidth}{c|c}
		Access Method  & Rate convergence time (s) \\
                \hline
                CRA = 50 kbps / RBDC = 1000 kbps  & 10  \\
                \hline
                CRA = 100 kbps / RBDC = 900 kbps  & 9  \\
                \hline
                CRA = 500 kbps / RBDC = 500 kbps  & 6  \\
                \hline
                CRA = 1000 kbps / RBDC = 0 kbps  & 7  \\
        \end{tabularx}
\end{table}

The results of this experiment are presented in
Table \ref{table:access_uncongested}. A low value of CRA impacts the end user
speed convergence time. That being said, once the threshold of 50\% of the
capacity on the return link is reached, increasing it beyond does not seem to
bring significant gains.

\subsection{Dynamic SATCOM access and congested network}

The previous test has shown that considering a low value of CRA, for example at
10\% of the capacity, increases the rate convergence time.
That being said, an actual system will likely be loaded. The test presented in
this section considers 9 UEs whose connection is established before the last UE
starts downloading.

The characteristics of this test scenario are as follows:
\begin{itemize}
	\item Number of connected UEs: 10;
	\item Same application for all UEs: data transfer in download;
	\item 9 UEs start downloading from the start, and one UE starts the download 10 seconds later;
	\item OpenSAND limits: 20 Mbps Forward / 1 Mbps in Return;
	\item Flow limitation per UE (SLA) in download: 2 Mbps;
	\item OpenSAND access (total return channel at 1 Mbps):
		\begin{itemize} 
			\item CRA = 50 kbps / RBDC = 1000 kbps
			\item CRA = 100kbps / RBDC = 900 kbps
			\item CRA = 500 kbps / RBDC = 500 kbps
			\item CRA = 1000 kbps /RBDC = 0 kbps
		\end{itemize}
\end{itemize}

The reported metric in this report is the rate convergence time, \textit{i.e.}
the time required for the 10$^{th}$ UE to reach the rate of its SLA. This
enables the analysis of the impact of the access mechanism
on the speed of convergence of congestion control.

\begin{table}[h]
	\caption{Rate convergence time for the 10$^{th}$ UE when all others UE started at the beginning of the connection}
\label{table:access_congested}
        \begin{tabularx}{\linewidth}{c|c}
		Access Method  & Rate convergence time (s) \\
                \hline
                CRA = 50 kbps / RBDC = 1000 kbps  & 4 \\
                \hline
                CRA = 100 kbps / RBDC = 900 kbps  & 11  \\
                \hline
                CRA = 500 kbps / RBDC = 500 kbps  & 10  \\
                \hline
                CRA = 1000 kbps / RBDC = 0 kbps  & 10  \\
        \end{tabularx}
\end{table}

The results of this experiment are presented in
Table \ref{table:access_congested}.  The result of the case 'CRA = 50 kbps /
RBDC = 1000 kbps' shows a very short convergence time. This may be due to the
link load which is variable. This phenomenon could have been absorbed by a
larger number of tests, which could not be carried out due to lack of time.
Moreover, the comparison of the other cases completes the results of the
previous section. When the CRA is set to a value greater than 100 kbps, the
convergence performance is the same. Once the network is loaded, it is not
necessary to dynamically adapt the use of the resource on the return path.

\subsection{Dynamic SATCOM access and mixed upload and download traffic}

In order to complete the results observed in the previous sections, tests with
mixed download and upload traffic were carried out and a subset of the results
is presented in this section.

The characteristics of this test scenario are as follows:
\begin{itemize}
	\item Number of connected UEs: 10;
	\item Application data transfer in download for 8 UEs and Upload for 2 UEs; 
		\begin{itemize} 
			\item Long flows for download and upload last 30 seconds;
			\item Short flows are 1 MB in download and 300 kB in upload; 
			\item 7 UEs start long flows (download) and 1 UE start long flows (upload) at the beginning of the experiment; 
			\item Short flows (1 UE in download and 1 UE in upload) start 10 seconds later; 
		\end{itemize}
	\item OpenSAND limits: 20 Mbps Forward / 1 Mbps in Return;
	\item Flow limitation per UE (SLA): 2 Mbps (download) and 300 kbps (upload);
	\item OpenSAND access (total return channel at 1 Mbps):
		\begin{itemize} 
			\item CRA = 50 kbps / RBDC = 1000 kbps
			\item CRA = 100kbps / RBDC = 900 kbps
			\item CRA = 500 kbps / RBDC = 500 kbps
			\item CRA = 1000 kbps /RBDC = 0 kbps
		\end{itemize}
\end{itemize}

\begin{table}[h]
	\caption{Short flows download and upload times in a congested environment}
\label{table:up-dw-acc}
        \begin{tabularx}{\linewidth}{c|c|c}
		Access Method  & 1 MB download time (s) & 300 kB upload time \\
                \hline
		CRA = 50 kbps  & 10.3 & 7.5 \\
		RBDC = 1000 kbps  &  &  \\
                \hline
		CRA = 100 kbps  & 10.7 & 7.2 \\
		RBDC = 900 kbps  & & \\
                \hline
		CRA = 500 kbps  & 9.9 & 7.1 \\
		RBDC = 500 kbps  & &  \\
                \hline
		CRA = 1000 kbps  & 11.6 & 7.5  \\
		RBDC = 0 kbps  &  &  \\
        \end{tabularx}
\end{table}

The results of this experiment are presented in Table~\ref{table:up-dw-acc}.
The different access mechanisms, once the network is loaded, do not seem to
illustrate substantial gains for the file sizes considered.

\subsection{Discussion}

The main conclusions of this activity are:
\begin{itemize}
	\item Once the network is loaded, 
		\begin{itemize}
			\item The impact of the choice of access method does not matter;
			\item CRA / RBDC or SCPC combinations (i.e. all capacity in CRA) show similar performance 
		\end{itemize}
	\item If the network is not loaded,
		\begin{itemize}
			\item A CRA value that is too low (lower than 10\% of maximum capacity) can impact performance;
			\item A CRA / RBDC approach can reduce costs.
		\end{itemize}
\end{itemize}

\section{Impact of transport protocol mechanisms}
\label{sec:access_channel}

This section provides the experiments that have led to assess the impact
of transport protocol mechanisms in a GEO-satellite backhaul system.

\subsection{Experiment set up}

This section presents the results for a web access or for a short file transfer.
The characteristics of the scenarios presented in this section are the following:
\begin{itemize}
	\item Number of connected UEs: 10;
	\item Test duration: 30 seconds;
	\item Data transfer is started for 9 UEs to load the link: 7 in download and and 2 in upload;
	\item The 10th UE consumes a given type of service (VoIP, video, Web or File transfer);
	\item The 9 UEs that load the link start their activity at the same time;
	\item The 10th UE that consumes the service starts a few seconds later, once the link is loaded;	
	\item OpenSAND limits: 20 Mbps Forward / 1 Mbps in Return;
	\item OpenSAND access: CRA = 100 kbps / RBDC = 900 kbps or CRA = 500 kbps / RBDC = 500 Kbps;
	\item Flow limitation per UE (SLA) of 2 Mbps in download and 100 kbps in upload.
\end{itemize}

The results presented in this section do not take into account the diversity of
		web pages and protocols used. For example, a page using the
		HTTP1 protocol and multiple objects has very different
		characteristics and probably different performance than a page
		using HTTP2.

The tests concerning the use of application traffic representative of a voice
		over IP or a video transmission did not show different results,
		whether the WAN accelerators are activated or not and whatever
		the configuration of the access (CRA = 100 kbps / RBDC = 900
		kbps, CRA = 500 kbps / RBDC = 500 kbps).

\subsection{Focus on web transfer}

In order to limit the complexity of the analysis, it was decided to test a
single web page with the following characteristics: 6.5 MB page size using the
HTTP protocol. On a 2 Mbps capacity link, the optimal transfer time would be
26 seconds.

\begin{table}[h]
\centering
\caption{Simple web page downloading time}
\label{table:pep_web}
	\begin{tabular}{c|c|c|c|c}
		Access Method  & \multicolumn{2}{c|}{CRA=100kbps} & \multicolumn{2}{c}{CRA=500 kbps} \\
		 & \multicolumn{2}{c|}{RBDC=900kbps} & \multicolumn{2}{c}{RBDC=500kbps} \\
		\cline{2-5}
		 & No PEP & PEP & No PEP & PEP \\
                \hline
		Average (s) & 33.9 & 36.1 & 34.5 & 34.8 \\
		Max (s) & 35.9 & 40.8 & 41.1 & 38.0 \\
		Min (s) & 31.0 & 32.2 & 31.6 & 32.4 \\
        \end{tabular}
\end{table}

The calculation of a statistic analysis based on 10 experiments is presented in
Table~\ref{table:pep_web}. The transfer times of the page vary between 31 and
41 seconds, all configurations combined. The introduction of a PEP does not
bring significant gain on web browsing in a congested context. It is worth
pointing out that these experiments do not consider the exploitation of a PEP
equipment where it should provide benefits, \textit{i.e.} within the SATCOM
system. This was not possible due to the lack of GTP-U capable PEPs. 

\subsection{Focus on file transfer}

The client proceeds to two different downloads during the experiment. During the
first one (fetch 1), the network is loaded with cross-traffic. The second download (fetch 2)
occurs when no other UE are using the satellite resource.  Each configuration
is tested five times.

\begin{table}[h]
\centering
\caption{Simple web page downloading time and CRA=100 kbps, RBDC=900 kbps}
\label{table:pep_file_100}
	\begin{tabular}{c|c|c|c|c}
		 & \multicolumn{2}{c|}{No PEP} & \multicolumn{2}{c}{PEP} \\
		\cline{2-5}
		 & Fetch 1 & Fetch 2 & Fetch 1 & Fetch 2 \\ 
		 Average download time (s) & 11.67 & 9.44 & 11.34 & 8.16 \\
        \end{tabular}
\end{table}

\begin{table}[h]
\centering
\caption{Simple web page downloading time and CRA=500 kbps, RBDC=500 kbps}
\label{table:pep_file_500}
	\begin{tabular}{c|c|c|c|c}
		  & \multicolumn{2}{c|}{No PEP} & \multicolumn{2}{c}{PEP} \\
		\cline{2-5}
		  & Fetch 1 & Fetch 2 & Fetch 1 & Fetch 2 \\ 
		 Average download time (s) & 11.91 & 9.32 & 11.49 & 6.88 \\
        \end{tabular}
\end{table}

Tables~\ref{table:pep_file_100} and~\ref{table:pep_file_500} present the
results for this experiment. In general, all fetchs 1 show the same values. It
means that whatever the channel access characteristics and whatever the
transport layer, when the network is loaded, the results are the same. However,
when the network is not loaded (fetch 2), including a PEP results in 3 \% to 26
\% performance improvements. The gains are more important when the CRA is high.

\subsection{Discussion}

Regarding file transfer, the gains brought by WAN acceleration are significant
without congestion, and amplify the conclusions on the adaptation of the access
method: the higher the CRA, the higher the gain brought by the PEP. Regarding web browsing, 
for a simple web page and in congestion, the WAN accelerator does not bring significant gains.

\section{Discussion}
\label{sec:discussion}

The main contributions of all these studies are as follows:
\begin{itemize}
	\item Different proofs of concept for the GEO backhaul service have been implemented;
	\item If the system is congested, the protocol optimizations offered by a PEP (to improve QoE) or adaptation of access mechanisms (to reduce costs) do not bring significant gains;
	\item If the system is not congested: 
		\begin{itemize}
			\item Protocol optimizations bring significant gains for file transfer;
			\item Adaptation of access mechanisms, \textit{i.e.} the constant reduction of the allocated throughput, enables to reduce the costs of access while having a negligible impact on services. 
		\end{itemize}
\end{itemize}
	
The tests carried out do not take into account the complexity of the operator's
core networks.  The introduction of WAN accelerator equipment (i.e. PEP)
ensures performance in the part for which the operator is responsible and
neglects the impacts of the network conditions between the operator's network and the data
servers. These equipments 
 isolate error segments, perform local retransmissions and tune the protocols to the network where it is deployed. 
They also implement caching mechanisms. 
Although negligible gains were measured by the introduction of these equipments,
this observation does not allow us to deduce its uselessness given that many functions 
were not considered and evaluated. 

The results nevertheless show the importance of studying the
impacts of the different protocol layers for SATCOM systems offering a
backhauling service for mobile networks. This includes many multi-layered
technical interactions, as well as the understanding of which is necessary 
to optimally size and implement such systems.

\section{Acknowledgments}
\label{sec:ackno}

This study was founded by CNES SMILE project. The authors would like to thank all those who participated in making this study possible.

\bibliographystyle{IEEEtran}
\bibliography{reference}

\end{document}